\documentclass[journal]{IEEEtran}

\usepackage{cite}
\usepackage{url}
\usepackage{bbm}
\usepackage{textcomp}
\usepackage{xcolor}
\usepackage{hyperref}

\usepackage{booktabs}	% 插表格用的宏包
\usepackage{diagbox}    % 插表格用的宏包
\usepackage{multirow}   % 插多行表格用的宏包

\usepackage{verbatim}	% 多行注释用的宏包

\usepackage{amsmath,amssymb,amsfonts,graphicx}
\usepackage{epstopdf}

\newtheorem{definition}{Definition}[section]
\newtheorem{property}{Property}[section]
\newtheorem{strategy}{Strategy}    % gws

\newtheorem{theorem}{Theorem}   % gws

\usepackage{graphicx,epstopdf,algpseudocode,caption,url}   % gws
\usepackage[ruled,linesnumbered]{algorithm2e}
\usepackage{amssymb}
\usepackage{hyperref}

\usepackage[switch]{lineno}   %gws 20170930
\begin{document}
% paper title

\title{Privacy-Preserving Federated Discovery of DNA Motifs with Differential
Privacy}

\author{Yao Chen, Wensheng Gan, Gengsen Huang, Yongdong Wu, Philip S. Yu,~\IEEEmembership{Life Fellow,~IEEE} 
		
%\thanks{Manuscript received Oct 31, 2021; revised XX 2022.}}

\thanks{This research was supported in part by the National Natural Science Foundation of China (Nos. 62002136 and 62272196), Natural Science Foundation of Guangdong Province (No. 2022A1515011861), Guangdong Key R\&D Plan2020 (No. 2020B0101090002), National Key R\&D Plan of China (No. 2020YFB1005600), Engineering Research Center of Trustworthy AI, Ministry of Education (Jinan University), and Guangdong Key Laboratory for Data Security and Privacy Preserving.}

\thanks{Yao Chen, Wensheng Gan, Gengsen Huang, and Yongdong Wu are with the College of Cyber Security, Jinan University, Guangzhou 510632, China. (E-mail: csyaochen@gmail.com, wsgan001@gmail.com, hgengsen@gmail.com, wuyd175@gmail.com)}
	
\thanks{Philip S. Yu is with the University of Illinois Chicago, Chicago, USA. (E-mail: psyu@uic.edu)} 

\thanks{Corresponding authors: Wensheng Gan and Yongdong Wu}
}

% make the title area
\maketitle

% As a general rule, do not put math, special symbols or citations
% in the abstract or keywords.

\begin{abstract}
    DNA motif discovery is an important issue in gene research, which aims to identify transcription factor binding sites (i.e., motifs) in DNA sequences to reveal the mechanisms that regulate gene expression. However, the phenomenon of data silos and the problem of privacy leakage have seriously hindered the development of DNA motif discovery. On the one hand, the phenomenon of data silos makes data collection difficult. On the other hand, the collection and use of DNA data become complicated and difficult because DNA is sensitive private information. In this context, how discovering DNA motifs under the premise of ensuring privacy and security and alleviating data silos has become a very important issue. Therefore, this paper proposes a novel method, namely DP-FLMD, to address this problem. Note that this is the first application of federated learning to the field of genetics research. The federated learning technique is used to solve the problem of data silos. It has the advantage of enabling multiple participants to train models together and providing privacy protection services. To address the challenges of federated learning in terms of communication costs, this paper applies a sampling method and a strategy for reducing communication costs to DP-FLMD. In addition, differential privacy, a privacy protection technique with rigorous mathematical proof, is also applied to DP-FLMD.  Experiments on the DNA datasets show that DP-FLMD has high mining accuracy and runtime efficiency, and the performance of the algorithm is affected by some parameters.
\end{abstract}

\begin{IEEEkeywords}
    DNA, motif discovery, federated learning, differential privacy, collaborative computing
\end{IEEEkeywords}

\IEEEpeerreviewmaketitle

\section{Introduction}  \label{sec:introduction}

In the era of big data, rich data not only enhances the core competitiveness of enterprises but also advances the progress of scientific research \cite{gan2017data,gan2021survey,gan2021fast,fecher2015drives}. The 5V characteristics of big data are volume, velocity, variety, value, and veracity \cite{nti2022mini}. However, these data often exist in the form of data silos, and there is a risk of privacy leakage. The phenomenon of data silos and the problem of privacy breaches hinder the flow and use of data. On the one hand, data silos hinder access to data. On the other hand, the issue of privacy leakage hinders the use of data \cite{gan2018privacy}. In this context, research fields involving private information are greatly hampered. DNA motif discovery is an important issue in gene research, which aims to identify transcription factor binding sites (i.e., motifs) of DNA sequences to reveal the mechanisms that regulate gene expression \cite{2007A}. A large amount of DNA information is a prerequisite for studying DNA motif discovery. However, DNA information contains private information about people, such as personal characteristics, diseases, and personality disorders. It has been demonstrated that the discovery of DNA motifs raises serious privacy concerns \cite{gymrek2013identifying}. Wu \textit{et al.} \cite{DBLP:journals/access/WuWSWX19}  showed that private information can be obtained by ordinary DNA motif discovery methods when no privacy protection methods are adopted. In addition, Homer \textit{et al.} \cite{2008Resolving} demonstrated that the specific identity of a person can be identified from a set of DNA data. Citizens and countries attach great importance to privacy protection. Countries around the world have introduced national data security laws and regulations, such as the GDPR in the EU \cite{voigt2017eu} and the CCPA in the USA \cite{pardau2018california}. These regulations greatly limit the misuse of private data and protect citizens' privacy rights. However, privacy protection can be an obstacle when it comes to the use of data that contains private information. For example, the National Institutes of Health withdrew all publicly available genomics data for clustering analysis. This has significantly limited the development of the biological sciences. Based on the above background,  how to conduct DNA motif discovery while solving data silos and ensuring privacy has been a very important task.

Achieving privacy protection in the process of DNA motif discovery is a challenging problem. Since the attacker may have some background knowledge, e.g., the attacker may already have information about all records except a specified one (i.e.,  the maximum background knowledge attack). In addition, the attacker's attack pattern is unknown. Differential privacy (DP) \cite{dwork2008differential} is a very suitable technique for solving this type of problem. DP mathematically proves that even if an attacker has the maximum background knowledge, it cannot determine the private data contained in this record \cite{wu2022adaptive}. The main idea of DP is not to provide privacy protection for the overall characteristics of the dataset, but to provide privacy protection for each individual in the dataset. DP can be used in DNA motif discovery to ensure that privacy is not compromised. The DNA motif discovery process is related to the problem of frequent sequential pattern mining \cite{agrawal1995mining}. However, the DP algorithm used for frequent pattern mining cannot be directly used in DNA motif discovery. This is because genes are subject to mutations, insertions, and deletions, which makes it difficult to ensure the appearance of motifs \cite{wu2019differential}. Chen \textit{et al.} \cite{chen2014private} devised a method for incorporating DP into the DNA motif discovery process. However, this method is inefficient and has many duplicate privacy budgets. To solve this problem, Xia \textit{et al.} \cite{xia2021secure} proposed a DNA motif discovery method based on sampling candidate pruning, in order to find DNA motifs with high practicability and efficiency. However, the application scenarios of these solutions are non-distributed. A solution suitable for distributed scenarios is very practical for dealing with big data.

Genetic research is based on a large amount of DNA data, which contains much private and valuable information. Research involving DNA data often encounters the problem of data silos. There are two main reasons. i) Medical institutions are unwilling to share patient information. The reason is that the importance attached to information security and the revenue mechanism of sharing medical data are not perfect. ii) The information systems of each medical institution are not unified, making it difficult to share information. Previously, the simplest solution to the problem of data silos was to integrate the data on one side and then perform further data processing. However, this method has many drawbacks, such as compromising privacy, requiring significant communication costs, and losing information. In such an environment, federated learning (FL) \cite{mcmahan2017communication} as an effective solution has attracted widespread attention. FL is a distributed machine learning technique that shows advantages in data utilization and multi-party training of models \cite{chen2023privacy}. The core idea of FL is that each participant builds a global model by sharing model parameters or intermediate results without sharing local individual or sample data. FL utilizes the data, computing power, and model-building capabilities of the participants to train the models. Therefore, the computational power and memory requirements of the server are reduced. Therefore, FL is a good technology that can be used to solve the data silo problem in different pattern discovery tasks \cite{li2022frequent,chen2023privacy}, especially DNA motif discovery.

At present, there are no studies examining how to apply FL to DNA motif discovery tasks. Therefore, how to use FL to solve the data silo problem in DNA motif discovery tasks is one of the challenges of the study. In addition, how to combine FL and DP is also a challenging problem. Finally, since there are multiple communications between participants and servers, reducing communication costs is also a concern. In this paper, we propose DP-FLMD, which applies FL and DP for solving privacy issues and achieving the goals of federal modeling in DNA motif discovery tasks. In addition, we put forward a communication reduction strategy to reduce communication costs. DP-FLMD is applied to a scenario where multiple participants jointly discover DNA motifs while protecting privacy. These participants are not willing to share their raw data with a third party and are only willing to provide some parameters to the third party. In the DP-FLMD framework, each participant performs a local differential privacy (LDP) method, which is used to add noise to the uploaded parameters. Therefore, the server cannot infer the participants' privacy from the noisy parameters. DP-FLMD uses a query-response approach between the server and the participants. The server sends a query and the participant sends a binary answer (i.e., a response) to the server. The server collects these answers and trains the model. Finally, the server obtains the desired DNA motifs. Our contribution is summarized as four points:

\begin{itemize}
    \item To the best of our knowledge, DP-FLMD is the first approach to use FL for DNA motif discovery. It solves both the problem of protecting privacy when finding DNA motifs and the problem of federated modeling.
    
    \item DP-FLMD uses a communication reduction strategy, which has two advantages. One advantage is that it reduces the number of messages sent by the server, thus reducing the communication cost required for FL modeling, and the other advantage is that it reduces the overall response time of the local client.
    
    \item DP-FLMD satisfies $\epsilon-$differential privacy by applying the randomized response algorithm. In addition, the randomized response method and the client sampling operation lead to randomness in the results. Therefore, to reduce the error caused by randomness, a threshold correction strategy is used in this paper to improve the utility of the data.
    
    \item Comprehensive experiments show that DP-FLMD can achieve high performance. In addition, we analyze the influence of parameters (e.g., different privacy levels, filtering error thresholds, and the number of participants) on DP-FLMD.
\end{itemize}

The rest of the article is structured as follows: In Section \ref{sec:prelim}, background and related work are presented. In Section \ref{sec:preliminary}, we provide the definition of DNA motif discovery, the definition of LDP, and the problem statement. In Section \ref{sec:algorithm}, we introduce the proposed algorithm, called DP-FLMD. Section \ref{sec:experiments} includes experimental results and analysis. Finally, conclusions and future work are presented in Section \ref{sec:conclusion}.

\section{Background and related work} \label{sec:prelim}

In this section, the background and related work of this paper are introduced. The background includes privacy-preserving DNA motif discovery, federated learning, and differential privacy.

\textbf{Privacy-preserving DNA motif discovery.} DNA motif refers to the transcription factor binding site of a DNA sequence \cite{2007A}. Since DNA sequences involve the user's private information, such as disease information, that is highly sensitive data. In order to protect the privacy of users, many countries have introduced laws and regulations for data security. Therefore, in the process of discovering DNA motifs, we must take the issue of data security and compliance seriously. There have been some meaningful research studies on private-preserving DNA motif discovery. Chen \textit{et al.} \cite{chen2014private} showed that DP can be used in the DNA motif discovery process and has the effect of protecting the privacy of DNA owners. However, privacy is only one aspect of the DNA motif discovery problem. The utility and accuracy of DNA motif discovery methods are also important. The DP-MFSC algorithm \cite{xia2021secure} was proposed in order to improve the privacy and utility trade-offs of privacy motif discovery algorithms. DP-MFSC provides higher utility and privacy by applying a sampling-based candidate pruning method, a sequence-length reduction method, and a threshold-modification strategy. To the best of our knowledge, there are no papers that consider both the problem of data silos and the issue of privacy protection in DNA motif discovery. Therefore, this paper aims to solve the problem of multiple participants jointly performing a private DNA motif discovery task.

\textbf{Federated learning.}  FL can effectively solve the problem of extracting valuable information from multiple datasets while maintaining data privacy. FL is a multi-party collaborative machine learning model with application scenarios for multiple participants to jointly train models with guaranteed data privacy \cite{chen2022federated}. In FL, the server performs three operations cyclically, i.e., collect, aggregate, and update, until the trained model reaches convergence. A round of model training can be divided into three steps. First, the server generates a model and then sends it to the participants. Then the participants train the model locally and send the trained model to the server. Finally, the server collects the models sent by the participants and then updates the models. The benefits of FL are improved data privacy, reduced communication costs for raw data transmission, and a reasonable trade-off between privacy and utility. FL has two challenges, namely high communication costs and privacy leakage issues. When the number of participants is large, although the quality of the model is improved, the communication overhead also increases. Communication costs can be reduced by reducing training rounds and the amount of information transmitted \cite{zhang2022challenges}. Zhu \textit{et al.} \cite{zhu2019deep} suggested that FL may suffer from gradient leakage, which can lead to privacy breaches. LDP can be used to solve the gradient leakage problem \cite{liu2020fedsel}. Many studies have applied FL to emerging applications in smart healthcare, such as EHRs management \cite{hao2020privacy}, federated medical imaging \cite{sheller2020federated}, federated remote health monitoring \cite{DBLP:conf/bionlp/LiuDM19}, and federated COVID-19 detection and diagnosis \cite{DBLP:journals/asc/FekiAKM21}. However, most of these applications belong to the field of data analysis. In the field of data mining, the question of how to combine FL and data mining to gain meaningful medical knowledge has not been thoroughly investigated.

\textbf{Differential privacy.} DP is a new privacy definition proposed by Dwork \textit{et al.} \cite{dwork2008differential} for the privacy leakage problem of statistical databases, which can be used to solve differential attacks. Under this privacy definition, the presence or absence of any one record in the dataset has negligible impact on the statistical results. Therefore, the privacy disclosure risk of a record caused by its being added to the dataset is controlled within an acceptable range. An attacker cannot accurately obtain the privacy information contained in this record by observing the statistical results. In addition, DP provides a rigorous definition and a quantitative evaluation method for the level of privacy protection. These advantages of DP make it a hot spot for privacy protection. DP can be used to solve the privacy leakage problem of frequent pattern mining \cite{chen2023privacy}. Xu \textit{et al.} \cite{xu2015differentially} first developed a DP-compliant FSM algorithm (named PFS2) to address the privacy leakage problem in the frequent sequence pattern mining process. However, PFS2 does not consider the difference in importance of different candidate sequences when the sequence database is reconstructed. The DP-FSM algorithm uses a heuristic approach to design scoring functions to distinguish the importance of candidate sequences \cite{DBLP:conf/icic/ZhouL18}. However, these algorithms are not efficient. Wang \textit{et al.} \cite{DBLP:journals/jksucis/WangH22} proposed a PrivFSM mechanism with high efficiency and high data utility. PrivFSM constructs prefix trees using sequences in the dataset under LDP, in order to mine frequent sequences.

\textbf{Differences}. In this paper, we study the problem of privacy leakage and data silos in DNA motif discovery. It is related to two studies. The first work \cite{xia2021secure} studies the problem of privacy leakage in the DNA motif discovery process. It proposes a DNA motif discovery algorithm with high usability and privacy, named DP-MFSC. It uses DP to protect privacy. To improve privacy and utility, this algorithm employs a sampling-based candidate pruning technique, a sequence length reduction method, and a threshold correction method. The second work \cite{wang2022fedfpm} investigates the problem of how to use FL to solve the problem of data silos in frequent pattern mining task. FedFPM the first algorithm to apply FL for frequent pattern mining, adopting the Apriori property to generate candidate patterns. FedFPM obtains support for candidate patterns by applying an interactive query response method between the server and the participants and then filters frequent patterns from the candidate patterns. In this paper, by using DP to protect privacy and FL, DP-FLMD can achieve the goal of joint modeling of multiple participants in the DNA motif discovery task. In addition, it further investigates how to reduce communication costs and improve data utility.

\section{Preliminary and problem statement} \label{sec:preliminary}
\subsection{DNA motif discovery}

The DNA motif discovery problem is to discover motifs that satisfy some constraints from a DNA sequence dataset. Given a DNA sequence dataset (denoted as $\mathcal{D}$). The user gives the length range of motifs (denoted as $[l_{min}$, $l_{max}]$), the support threshold (denoted as $f$), the error tolerance threshold (denoted as $\delta$), and the number of desired motifs (denoted as $N$). The DNA motif discovery problem is to discover frequent sequences in $\mathcal{D}$ with lengths within $[l_{min}$, $l_{max}]$ and the value of consolidated frequency ranked in the top $N$.

\begin{definition}
  \rm (Frequent sequence \cite{pei2001mining}). A frequent sequence is defined as a sequence in a dataset (denoted as $\mathcal{D}$) whose support is not less than a user-given support threshold (denoted as $f$). The support of a sequence (denoted as $p$) is defined as:
\begin{equation}\label{eq1}
	sup(p,\mathcal{D}) = \frac{count(p,\mathcal{D})}{|\mathcal{D}|},
\end{equation}
where $count(p,\mathcal{D})$ denotes the number of occurrences of $p$ in $\mathcal{D}$ and $|\mathcal{D}|$ denotes the number of transactions in $\mathcal{D}$.
\end{definition}

\begin{definition}
    \rm (Hamming distance \cite{hamming1950error}). Suppose there are two sequences of the same length $S$ = $ \langle s_0,s_1,...,s_{l-1} \rangle$ and $P$ = $ \langle p_0,p_1,$...$,p_{l-1} \rangle$. The Hamming distance between $S$ and $P$ is defined as:
\begin{equation}\label{eq2}
    d_H(S,P) = |\{i|s_i \oplus p_i = 1, 1 \leq i \leq l\}|.
\end{equation}
\end{definition}

For example, $S$ = $\langle AGTCA \rangle$ and $P$ = $\langle ATTCG \rangle$. According to the above equation, we can derive $d_H(S,P)$ = 2.

\begin{definition}
     \rm (Approximate motifs \cite{ferreira2006mining}). When two sequences have the same length and the Hamming distance is less than a user-specified error tolerance (denoted as $\delta$), they are considered approximate motifs.
\end{definition}

For example, suppose $\delta$ = 3, $S$ = $\langle AGTCA \rangle$, and $P$ = $\langle ATTCG \rangle$. Since $d_H(S,P)$ = 2 is less than $\delta$, they are approximate motifs.

\begin{definition}
    \rm (Consolidated frequency \cite{chen2014private}).  The consolidated frequency of sequence $S$ is defined as:
\begin{equation}\label{eq3}
\textit{cf}(S) = sup(S,\mathcal{D}) + \sum_{P \in \textit{MS}}sup(P,\mathcal{D}),
\end{equation}
where $sup(S,\mathcal{D})$ and $sup(P,\mathcal{D})$ denote the support of sequence $S$ and sequence $P$, respectively, and \textit{MS} denotes the set of approximate motifs of $S$.
\end{definition}

\subsection{Local differential privacy}

This paper uses DP to protect data privacy. DP proves through a rigorous mathematical derivation that even if an attacker has acquired all the records except for a specified record, it cannot determine the information contained in this record. The important definitions of DP are as follows:

\begin{definition} \label{ND}
    \rm (Neighboring datasets \cite{li2016differential}). Suppose there are two datasets $D_1$ and $D_2$. If $D_1$ and $D_2$ differ by only one record, they are considered neighboring datasets.
\end{definition}

\begin{definition} \label{DP}
    \rm ($\epsilon-$differential privacy, denoted as $\epsilon-$DP \cite{dwork2006differential}). Suppose there are neighboring datasets $D_1$ and $D_2$. If a randomized algorithm $\mathcal{M}$ satisfies Formula \eqref{eq4}, then $\mathcal{M}$ satisfies $\epsilon-$DP.
\begin{equation}\label{eq4}
          \forall y \in  Range(\mathcal{M}): 
        Pr[\mathcal{M}(D_1) =y] \leq e^{\epsilon} \times Pr[\mathcal{M}(D_2) =y],           
\end{equation}
where $Range(\mathcal{M})$ denotes the set of all possible outputs of $\mathcal{M}$. $\epsilon$ is called the privacy budget. When $\epsilon$ is smaller, it means that the level of privacy protection is higher.
\end{definition}

\begin{definition} \label{RR}
    \rm (Randomized response \cite{warner1965randomized}). The randomized response algorithm $\mathcal{M}$ is the most typical DP algorithm for perturbing the value of an input bit $v$. The purpose of $\mathcal{M}$ is to make the value of $v$ uncertain. Specifically, $v$ is the input to the algorithm $\mathcal{M}$, and then the output $y$ is obtained by executing the algorithm $\mathcal{M}$. $y$ exists in two cases, i.e., $y$ is equal to $v$ or $y$ is not equal to $v$. We define the probability that $y$ is not equal to $v$ as $\eta$. $\eta$ is called the noise factor and $\eta \in$ [0, 0.5).
\begin{equation}\label{eq5}
	\forall y \in Range(\mathcal{M}):
	Pr[\mathcal{M}(v)=y] =  \left\{ \begin{array}{ll}
		1 - \eta  ,& {\rm if} y= v\\
		\eta ,&  {\rm if} y \neq v
	\end{array} \right. 
\end{equation}
\end{definition} 

\begin{theorem}\rm \label{theorem:1}
    Randomized response algorithm $\mathcal{M}$ satisfies $\epsilon$-LDP when \begin{equation}\label{eq6}
    \eta = \frac{1}{1 + e^\epsilon}.
    \end{equation}
\emph{Proof.} The theorem has been proved by Wang \textit{et al.} \cite{wang2022fedfpm}. See Appendix A for proof details. 
\end{theorem}

\subsection{Problem statement}

In the DNA motif discovery problem, since DNA sequences contain a lot of private information about personal characteristics and diseases, each client does not want to share their original data. In this case, if we want to use these clients' data to train the model or obtain useful information, we cannot use the traditional approach to train the model by aggregating the clients' data with a third party. The original data for each client can only be stored locally. Each client shares only a few parameters with the third party (e.g., server) that do not compromise privacy. In short, the problem of this paper is how to federate multiple clients to discover motifs while protecting their privacy.

\section{Proposed Methods} \label{sec:algorithm}

In this section, we first introduce the definitions, properties, and strategies related to DP-FLMD. Then the flow and pseudocode of DP-FLMD are described. Finally, two important operations in DP-FLMD are introduced, including the generation of candidate patterns and the calculation of the consolidated frequencies.

\begin{definition}\label{PreSuf}
    \rm (Prefix and  Suffix \cite{han2001prefixspan, mccreight1976space}). Given a sequence $S$ = $ \langle s_0,s_1,...,s_{l-1} \rangle$. Suppose an element of $S$ is $s_i$. The prefix of $s_i$ is the subsequence from the beginning to $s_i$ in $S$, denoted as $Pre_i(S)$. The suffix of $s_i$ is the subsequence from $s_i$ to the ending in $S$, denoted as $Suf_i(S)$.
\end{definition}

\begin{definition}\label{MS}
    \rm (Merged sequence set). A merged sequence set is defined as a set that stores merged sequences, where each merged sequence can be split into multiple sub-sequences of the same length that differ only in the last character.
\end{definition}

For example, given a sequence $S$ = $\langle ABCDE \rangle $, $Pre_0(S)$ = $\langle A \rangle $, $Pre_2(S)$ = $ \langle ABC \rangle $, $Suf_0(S)$ = $\langle ABCDE \rangle $, and $Suf_2(S)$ = $\langle CDE \rangle $. Given a merged sequence set, $\mathcal{MS}$ = \{$\langle ABCABC \rangle$, $\langle ABCDAB \rangle$\}. Take the first merged sequence $\langle ABCABC \rangle$ as an example. The merged sequence can be divided into three sub-sequences of length 4, namely $\langle ABCA \rangle$, $\langle ABCB \rangle$, and $\langle ABCC \rangle$. These three sequences differ only in the last character.

\begin{property}
    \label{Apriori}
    \rm (Apriori property \cite{agrawal1994fast}). If the itemset $X$ is a frequent itemset, then all its non-empty subsets are frequent itemsets. In addition, if the itemset $X$ is not a frequent itemset, then none of its supersets are frequent itemsets.
\end{property}

\begin{strategy}
    \label{ICS}
    \rm (Communication reduction strategy). 
    The server sends all candidate patterns to the participants in order to obtain noisy responses. However, the number of candidate patterns generated by the central server is large. Sending these messages consumes communication resources significantly. To reduce the communication cost, the amount of information needs to be compressed. Suppose there are $k$ sequences with the same length (denoted as $l$). If they have the same $l-1$ prefix, then these $k$ sequences can be combined into a merged sequence. The length of the merged sequence is $l-1+k$. In addition to this, participants can know that these $k$ sequences with the same $l-1$ prefix do not exist locally by finding that $l-1$ prefix is not present in the local data. This strategy reduces the time spent by participants for querying candidate patterns. Therefore, the communication reduction strategy reduces the response time of the participants.
\end{strategy}

For example, suppose there are three sequences of length 4 with the same $l-1$ prefix, $S_0$ = $\langle ABCA \rangle$, $S_1$ = $\langle ABCB \rangle$, and $S_2$ = $\langle ABCC \rangle$. The three sequences are compressed into one using the communication reduction strategy, and the compressed sequence is $S$ = $\langle ABCABC \rangle$.

\begin{strategy}
    \label{MTS}
\rm (Threshold correction strategy). Due to the client sampling operation and the randomized response method, the mining results are unknown and random. We cannot claim with 100\% confidence that a candidate pattern (denoted as $p$) is a frequent pattern. Therefore, we use a modified threshold (denoted as $\theta$) to determine whether $p$ is a frequent pattern or not. The modified threshold introduces a parameter $\xi$, and $\xi$ is used to control the minimum confidence required for the decision.
	\begin{equation}\label{eq7}
	\theta =  f + \eta -2f \eta  + \sqrt{ \frac{- ln\xi}{2x} },
	\end{equation}
where $f$ denotes the frequent threshold given by the user, $\eta$ denotes the noise factor, which is calculated according to Theorem \ref{theorem:1}, and $x$ denotes the number of participants. $x$ also indicates the number of noisy responses received by the server about the candidate pattern $p$. Notice that the candidate $p$ will be a frequent pattern with  1-$\xi$ confidence when
\begin{equation}\label{eqtheorem2}
     \frac{n_1'}{x} \geq \theta,
\end{equation}
where $n_1'$ is the number of noisy responses with a value of 1. The derivation process of the formula for calculating the modified threshold $\theta$ (i.e., Equation \eqref{eq7}) is presented in Appendix B. In addition, the strategy has been proposed by Wang \textit{et al.} \cite{wang2022fedfpm}. 
\end{strategy}

\subsection{DP-FLMD algorithm}

\begin{figure}[!t]
\centering
\includegraphics[width=3in]{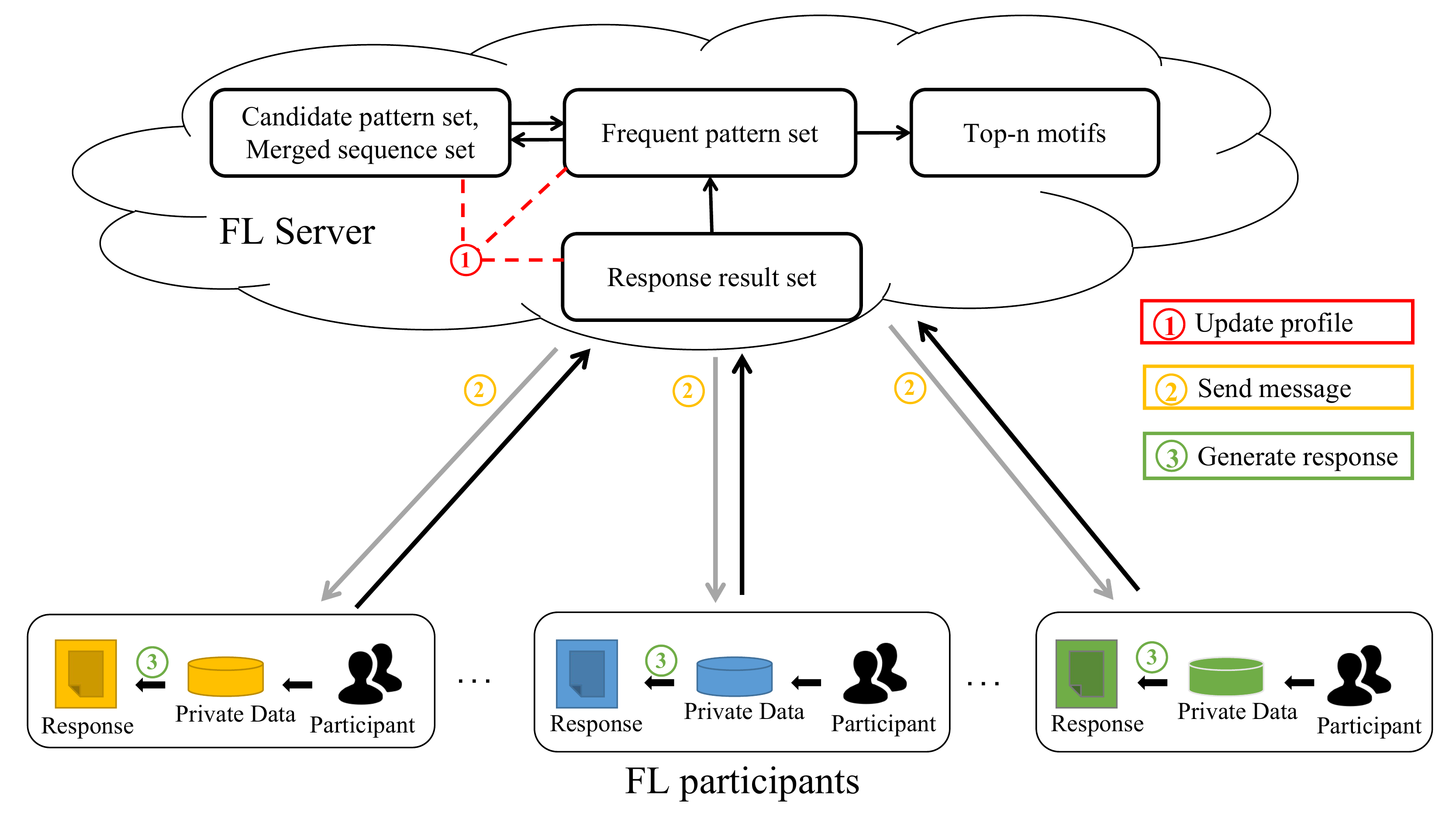}
\caption{An overview of DP-FLMD.}
\label{fig:overview}
\end{figure}

DP-FLMD can be divided into three steps: the server initializes the profile, the server and clients train together, and the server obtains results based on the profile. Algorithm \ref{alg:DP-FLMD} describes this workflow, and the details of Algorithm \ref{alg:DP-FLMD} are explained below. The first step is that the server initializes the profile and calculates the noise factor $\eta$ (lines 1-2). The profile consists of a candidate pattern set $C$, a frequent pattern set $F$, and a merged sequence set $\mathcal{MC}$. The second step is to perform multiple training (lines 3-29). The purpose of each training is to obtain the length $l$ of frequent patterns and candidate patterns. The last step is to obtain a set that stores the top $N$ motifs with the highest consolidated frequency (denoted as \textit{NCFM}) based on Algorithm \ref{alg:CCF} by calculating the consolidated frequency of the frequent patterns, where \textit{NCFM} is a set that stores the top $N$ motifs with the highest consolidated frequency (line 30). As shown in Fig. \ref{fig:overview}, each training consists of the following three key steps.

\begin{algorithm}[h]
	\small
	\caption{DP-FLMD algorithm}
	\label{alg:DP-FLMD}
	\LinesNumbered
	\KwIn{Data for clients: $\mathcal{D}$, set of clients: $\mathcal{N}$, number of participants in each round: $x$, threshold: $\theta$, error tolerance parameter: $\delta$, motif length range: $[l_{min}$, $l_{max}]$, the value $N$ in top-$N$, the parameter of LDP: $\epsilon$, filtering error threshold: $\xi$, alphabet: $A$.} 
	\KwOut{\textit{NCFM}.}
 
     $C$, $F$, $\mathcal{MC}$  = $\emptyset$\;
     $\eta$ $\leftarrow$ substitute $\delta$ into Formula \eqref{theorem:1} to calculate the noise factor $\eta$\;
	\For{\rm $l$ = $l_{min}$ to $l_{max}$}{
        \If{$l$ == $l_{min}$}{
        $F_{l-1}$ = $\emptyset$\;
        }
        $C,\mathcal{MC}$ = \textbf{GenerateCandidates}($F_{l-1}$, $l_{min}$, $l$, $A$)\;
        $R$ = $\emptyset$ $\qquad \rhd$ $R$ denotes the noisy response result set\;
        $\mathcal{PS}$ $\leftarrow$ select $x$ clients from $\mathcal{N}$ as the set of participants\;
        
        \For{\rm each $i$ $\in$ $\mathcal{PS}$}{
        \For{\rm each $m$ $\in$ $\mathcal{MC}$}{
                  \If{$Pre_{l-1}(m)$ $\in$ $\mathcal{D}[i]$}{
                      \textit{pos} $\leftarrow$ records the first occurrence of the last character of $Pre_{l-1} (m)$ in $\mathcal{D}[i]$\;
                     
                      \For{\rm each $s$ $\in$ $Suf_{l}(m)$}{                         
                           $r'$ = \textbf{ParticipantResponse}($Suf_{pos+1}$ $(\mathcal{D}[i])$, $s$, $\eta$)\;
                           $R$ $\leftarrow$ update $R$ according to $r'$\;
                     }
                   }
                   \Else{
                     \For{\rm $d=1$ to $Suf_{l}(m)$}{
                           $r = 0$\;
                           $r'$ $\leftarrow$ randomize $r$ with Formula \eqref{eq5}\;
                           $R$ $\leftarrow$ update $R$ according to $r'$\;
                     }
                   }
        }
     }
     $F$ $\leftarrow$ update $F$ according to $R$ and Formula \eqref{eqtheorem2}\;
     }
    \textit{NCFM} = \textbf{CalculateConsolidataFrequencies}($F$, $l_{min}$, $l_{max}$, $\delta$)\;
    
    \textbf{return} \textit{NCFM}
\end{algorithm}

\textbf{Step 1}: Update profile (lines 4-8, 28). In this step, the server updates the profile, which includes $F$, $C$, $\mathcal{MC}$, and $R$. $F_{l-1}$ is a frequent pattern set, which is obtained based on  the results of the last round of training, and the length of each frequent pattern in $F_{l-1}$ is $l-1$. Therefore, when this round is the first round, $F_{l-1} $ is an empty set (lines 4-6). At the beginning of each training, the server generates the candidate pattern set $C$ and the merged sequence set $\mathcal{MC}$ based on Algorithm \ref{alg:GC} (line 7). $\mathcal{MC}$ is the message sent by the server to the participant. Then the server initializes the noisy response result set $R$ (line 8). At the end of each training, the server updates the frequent pattern set $F$ based on $R$ collected in this round (line 28). 

\textbf{Step 2}: Send messages (line 9). The server selects participants from the clients (line 9). These participants are given the information $\mathcal{MC}$ sent by the server.

\textbf{Step 3}: Generate responses (lines 10-27). In this step, each participant $i$ obtains candidate patterns by decomposing the merged sequences in $\mathcal{MC}$, and then updates the noisy response set $R$ based on the local data and the randomized response method. According to Strategy \ref{ICS} and Definition \ref{MS}, a merged sequence $m$ stores the information of several candidate patterns of length $l$. These candidate patterns have the same $l-1$-prefix, namely $Pre_{l-1}(m)$. If $Pre_{l-1}(m)$ exists in the data of this participant (i.e., $\mathcal{D}[i]$), then these candidate patterns may also exist in $\mathcal{D}[i]$ (lines 12-18). Therefore, to determine whether a candidate pattern exists in $\mathcal{D}[i]$, it is only necessary to determine whether the last item of the candidate pattern exists in $Suf_{pos+1}(\mathcal{D}[i])$. Algorithm \ref{alg:PR} is used to obtain a noisy response (line 15). In Algorithm \ref{alg:PR}, the participant first obtains the original response $r$ by determining whether the last item $s$ is present in $Suf_{pos+1}(\mathcal{D}[i])$ or not. Then, the noisy response is obtained according to the definition of the randomized response (Definition \ref{RR}). The participant updates the noisy response result set $R$ based on the value of $r'$ (line 16). If $Pre_{l-1}(m)$ does not exist in the data of this participant (i.e., $\mathcal{\mathcal{D}}[i]$), then it means that these candidate patterns cannot exist in $\mathcal{\mathcal{D}}[i]$ (line 19-25). This means that the original response is 0 (line 21). The participant then performs the randomized response operation (line 22) and updates $R$ (line 23).

\subsection{Generation of candidate patterns}

The sub-sequences of frequent sequences must also be frequent sequences. Therefore, Property \ref{Apriori} can also be used in frequent sequence mining and be used to generate candidate patterns. Algorithm \ref{alg:GC} uses Property \ref{Apriori} to generate candidate patterns. The details of Algorithm \ref{alg:GC} are explained below.

The first step is that the server initializes the candidate pattern set $C$ and generates a string \textit{SA}, which contains all the characters of $A$ (lines 1-2). The second step is to determine whether this is the first round of training (lines 3-5). The value of $l$ is $l_{min}$ implying that this is the first round of training (line 3). This also means that $F_{l-1}$ is an empty set. Therefore, a value needs to be assigned to $F_{l-1}$. $F_{l-1}$ stores all the patterns of length \textit{l-1}, which are generated based on the elements of $A$ (line 4). The last step is to generate $C$ and $\mathcal{MC}$ based on $F_{l-1}$ (lines 6-13). The server traverses every frequent pattern of length \textit{l-1}. Based on the definition of the merged sequence set (Definition \ref{MS}), we know that the merged sequence is $p$ + \textit{SA} and place it in the merged sequence set $\mathcal{MC}$ (lines 7-8). Finally, according to the Apriori property, we generate candidate patterns of length $l$ (lines 9-12).

\begin{algorithm}[ht]
	\small
	\caption{GenerateCandidates}
	\label{alg:GC}
	\LinesNumbered
	\KwIn{The set of frequent sequences of sequence with the length \textit{l-1}: $F_{l-1}$, a minimum length of motif: $l_{min}$, the motif length: $l$, alphabet: $A$.} 
	\KwOut{The set of candidate patterns: $C$, the set of merged sequences: $\mathcal{MC}$.}
    $C$ = $\emptyset$\;
    \textit{SA} $\leftarrow$ connect all elements of $A$ $\qquad \rhd$ \textit{SA} is a string containing all the characters of $A$\;
    \If{$l$ == $l_{min}$}{
            $F_{l-1}$  $\leftarrow$ generate all patterns of length $l_{min-1}$ using elements in $A$\;
     }
    \For{\rm each $p$ $\in$ $F_{l-1}$}{
        $m$ = $p$ + \textit{SA} $\qquad \rhd$ $m$ is a merge pattern with length $l$-1+len($A$)\;
        add $m$ to $\mathcal{MC}$\;
        \For{\rm each $i$ $\in$ $A$}{
            $c$ = $p+i$ $\qquad \rhd$ $c$ is a candidate pattern with length $l$\;
            add $c$ to $C$\;
        }
    }
    \textbf{return} $C$, $\mathcal{MC}$
\end{algorithm}

\begin{algorithm}[ht]
	\small
	\caption{ParticipantResponse}
	\label{alg:PR}
	\LinesNumbered
	\KwIn{Local data: $d$, pattern: $i$, the noise factor: $\eta$.} 
	\KwOut{Response: $r'$.}
    $r$ $\leftarrow$  $\mathbbm{I}$($i$ $\subseteq$ $d$) $\qquad \rhd$ $\mathbbm{I}$ is called the indicator function to determine whether $i$ belongs to $d$\;
    $r'$ $\leftarrow$ randomize $r$ with Formula \eqref{eq5}\;
    \textbf{return} $r'$
\end{algorithm}

\subsection{Calculation of consolidated frequencies}

Motifs appear in the form of degradation, so in addition to the frequency of the motif itself, the frequency of the approximate motifs also needs to be considered. In other words, it needs to sort and filter the motifs according to their consolidated frequency. The function of Algorithm \ref{alg:CCF} is to calculate the consolidated frequency of all motifs of length in [$l_{min}$, $l_{max}$] and get the top $N$ motifs with the highest consolidated frequency. The details of Algorithm \ref{alg:CCF} are explained below. 

\begin{algorithm}[h]
	\small
	\caption{CalculateConsolidataFrequencies}
	\label{alg:CCF}
	\LinesNumbered
	\KwIn{The set of frequent sequences: $F$, the minimum length of motif: $l_{min}$, the maximum length of motif: $l_{max}$, error  tolerance parameter: $\delta$.} 
	\KwOut{A set that stores the top $N$ motifs with the highest consolidated frequency: \textit{NCFM}.}
    \textit{NCFM}, \textit{LMCS}  = $\emptyset$\;
    % Step 1: put all l-motifs into different sets.
    
    \textit{LMS} $\leftarrow$ get \textit{LMS} by traversing $F$ $\qquad \rhd$ \textit{LMS} stores motifs of each length\;
 
    % Step 2.
    \For{\rm $l$ = $l_{min}$ to $l_{max}$}{
    \If{\rm $l$  $\in$ \textit{LMS}}{
    % \If{\rm len(\textit{LMS}[$l$]) != 0}{
    \textit{seq}$\_l$ = \textit{LMS}[$l$]\;
    $s$ = \textit{seq}$\_l$.keys()[0]\;

    % Step 2.1：get Bucket
    \textit{Bucket} $\leftarrow$ get \textit{Bucket} by traversing \textit{seq}$\_l$\;
    
    % Step 2.2
    \For{\rm each $i$ $\in$ \textit{Bucket}.keys()}{
    \For{\rm each $p1$ $\in$ \textit{Bucket}[$i$]}{
    \textit{LMCS}.setdefault($l$, \{\})[$p1$] = 0 
    $\qquad \rhd$ The \textit{setdefault} method in python is used here to set the corresponding value of the specified key\;
    
    \If{\rm $i$  $\geq$ $\delta$}{
    \For{\rm $j$ = int($i$ - $\delta$) to min(($i$ +  $\delta$), $l$) + 1}{
       call \textbf{UpdateValue}($j$, \textit{Bucket}, $p1$, $l$, \textit{LMCS}, $\delta$)\;
    }
    }
    \Else{
    \For{\rm $j$ = 0 to min(int($i$ + $\delta$), 1) + 1}{
        call \textbf{UpdateValue}($j$, \textit{Bucket}, $p1$, $l$, \textit{LMCS}, $\delta$)\;
    }
    }
    }
    }
    $M$ = \textit{NCFM} + \textit{LMCS}[$l$]\;
    \textit{NCFM} $\leftarrow$ stores the top $N$ motifs with the highest consolidated frequency in $M$\;
    }
    }
    % Step 3.
    \textbf{return} \textit{NCFM}
\end{algorithm}

First, the server performs the initialization operation, i.e. \textit{NCFM} and \textit{LMCS} are set to the empty set (line 1). Then, the server puts all frequent patterns into different sets according to pattern length (line 2). For example,  \textit{LMS}[$l$] stores frequent patterns of length $l$. Finally, the server calculates the consolidated frequency for motifs of each length and updates \textit{NCFM}, where \textit{NCFM} is a set that stores the top $N$ motifs with the highest consolidated frequency (lines 3-26). If \textit{LMS} has frequent patterns of length $l$, then the consolidated frequency of  frequent patterns of length $l$ is calculated and \textit{NCFM} is updated (lines 4-25). The implementation can be divided into four steps.

\begin{algorithm}[ht]
	\small
	\caption{UpdateValue}
	\label{alg:Update}
	\LinesNumbered
	\KwIn{Location of the approximate motif: $j$, a collection that stores motifs: \textit{Bucket},  a motif: $p1$, the length: $l$, tolerance parameter: $\delta$.} 
	\KwOut{A set that stores motifs and their consolidated frequency for each length: \textit{LMCS}.}
    \If{\rm $j$ $\in$ \textit{Bucket}.keys()}{
    \For{\rm $p2$ $\in$ \textit{Bucket}[$j$]}{
     \If{\rm Levenshtein.hamming(p1, p2) $\leq$ $\delta$}{
     \textit{LMCS}.setdefault($l$, \{\})[$p1$] = \textit{LMCS}.setdefault($l$, \{\})[$p1$] + \textit{LMS}[$l$][$p2$]\;
     }
    }
    }
    \textbf{return} \textit{LMCS}
\end{algorithm}

The first step is to obtain \textit{seq}$\_l$ and $s$ (lines 5-6). \textit{seq}$\_l$ is used to store motifs of length $l$ (line 5). Then, select a reference motif (denoted as $s$) from \textit{seq}$\_l$. Here, the first element of \textit{seq}$\_l$ is selected as $s$ (line 6). The second step is to calculate the Hamming distance between $s$ and other motifs in \textit{seq}$\_l$ and classify these motifs according to their Hamming distances (line 7). Here, \textit{Bucket} is used to store these classified motifs. \textit{Bucket} is a dictionary that stores motifs with various Hamming distances. For example, \textit{Bucket}[$i$] stores every motif that satisfies the condition: the Hamming distance between it and $s$ is $i$. The third step is to obtain \textit{LMCS}[$l$] by traversing the elements of \textit{Bucket} (lines 8-22). \textit{LMCS}[$l$] stores motifs of length $l$ and their consolidated frequency. All keys of the dictionary \textit{Bucket} are traversed (line 8). Each motif in \textit{Bucket}[$i$] is traversed, and the Hamming distance between each motif and $s$ is $i$ (line 9). The consolidated frequency of $p1$ of length $l$ is set to 0 (line 10). There are two cases. The first case is that the Hamming distance $i$ is not less than $\delta$ (lines 11-15). The second case is that the Hamming distance $i$ is less than $\delta$ (lines 16-20). $j$ is used to determine the range of the approximate motif of $p1$. Algorithm \ref{alg:Update} is executed (lines 13 and 18). The purpose of Algorithm \ref{alg:Update} is to find approximate motifs of $p1$ and update \textit{LMCS}[$l$]. The approximate motifs of $p1$ exist in \textit{Bucket}[$j$], so it traverses all motifs in \textit{Bucket}[$j$] and determines whether the Hamming distance between each motif $p2$ and $p1$ is no greater than $\delta$. If so, the consolidated frequency of $p1$ needs to be updated, i.e., the noisy support of $p2$ is added to the consolidated frequency of $p1$. The final step of Algorithm \ref{alg:CCF} is to update \textit{NCFM} (lines 23-24).

\section{Experimental Evaluation} \label{sec:experiments}

In this section, we design several experiments to analyze the performance of DP-FLMD, which is used to solve the problem of multiple participants jointly mining motifs while protecting privacy. In addition, we investigate the effect of parameters on DP-FLMD, including the filtering error threshold, the parameter of LDP, and the number of participants in each round.

\subsection{Experimental settings}

\textbf{Implementation}. A large number of experiments were performed on different datasets. All experiments were conducted on a Macbook Air with an M1 chip and 16 GB of RAM. The algorithms involved in the experiments were implemented in Python. Due to the client sampling operation and the randomized response method, the mining results are unknown and random. Therefore, the value of each data point in experimental plots is the average of the results obtained by executing the algorithm 100 times.

\textbf{Datasets}. We used six DNA datasets, including: promoters \footnote{https://archive.ics.uci.edu/ml/machine-learning-databases/molecular-biology/promoter-gene-sequences.}, washington \footnote{http://bio.cs.washington.edu/assessment/download.html.}, chrUn \footnote{https://hgdownload.soe.ucsc.edu/goldenPath/mm39/chromosomes/chr\\Un\_MU069435v1.fa.gz}, splice \footnote{https://archive.ics.uci.edu/ml/machine-learning-databases/molecular-biology/splice-junction-gene-sequences.}, chrY \footnote{https://hgdownload.soe.ucsc.edu/goldenPath/mm39/chromosomes/chr\\Y\_JH584301v1\_random.fa.gz}, and centers \footnote{https://github.com/microsoft/clustered-nanopore-reads-dataset.}. Promoters and splice are two datasets from the UCI, where promoters is an E. coli promoter gene sequence dataset and splice is a dataset that stores splice-junction-gene sequences. Washington is a real DNA dataset. ChrUn and Chry are two datasets involving chromosome information in mouse. The Centers dataset contains 10,000 strings of length 110 that are randomly generated.

Table \ref{number} describes the characteristics of each dataset, including the number of transactions (denoted as $|\mathcal{D}|$) and the average length of a transaction (denoted as \textit{avelen}). A transaction of the dataset is considered a client's data. The table also provides the constraints that motifs need to meet, including the number of desired motifs (denoted as $N$), the motif length range (denoted as $[l_{min}$, $l_{max}]$), and the tolerance threshold of the motif (denoted as $\delta$). DP-FLMD discovers motifs that meet the above constraints on each dataset. In addition, the default values of the experiment parameters include the filtering error threshold (denoted as $\xi$), the parameter of LDP (denoted as $\epsilon$), and the number of participants in each round (denoted as $x$). When analyzing the impact of one parameter on the algorithm's performance, the other parameters are set to the default values. In Table \ref{number}, the datasets are sorted according to $|\mathcal{D}|$. Promoters with the smallest $|\mathcal{D}|$ is the first dataset in Table \ref{number}. In this paper, we assume that a larger value of $|\mathcal{D}|$ indicates a larger problem size.

\begin{table}[ht]
    \fontsize{5.5pt}{15pt}\selectfont
	\centering
	\caption{Characteristics and default parameters in datasets}
	\label{number}
	\begin{tabular}{|c|lllllllll|}
		%\toprule
		\hline\hline \textbf{Dataset} & $|\mathcal{D}|$ & \textit{avelen} & $N$ & $l_{min}$  & $l_{max}$  & $\delta$  & $\xi$   & $\epsilon$ & $x$ \\ \hline	
        Promoters & 106 & 57.0 & 30 & 1 & 4  & 1 & 0.01 & 3 & 53\\
        Washington & 522 & 16.7 & 30 & 1 & 3  & 1 & 0.01 & 3 & 261 \\
        ChrUn & 624 & 49.97 & 30 & 1 & 4  & 1 & 0.01 & 3 & 312\\
        Splice & 3190 & 60.0 & 30 & 1 & 5  & 1 & 0.01 & 3 & 1595\\
        ChrY & 5198 & 49.995 & 30 & 1 & 4  & 1 & 0.01 & 3 & 2599\\
        Centers & 10000 & 110.0 & 30 & 1 & 5  & 1 & 0.01 & 3 & 5000\\	
			\hline \hline		
			%\bottomrule
	\end{tabular}
\end{table}

\begin{figure}[!t]
\centering
\includegraphics[width=2.5in]{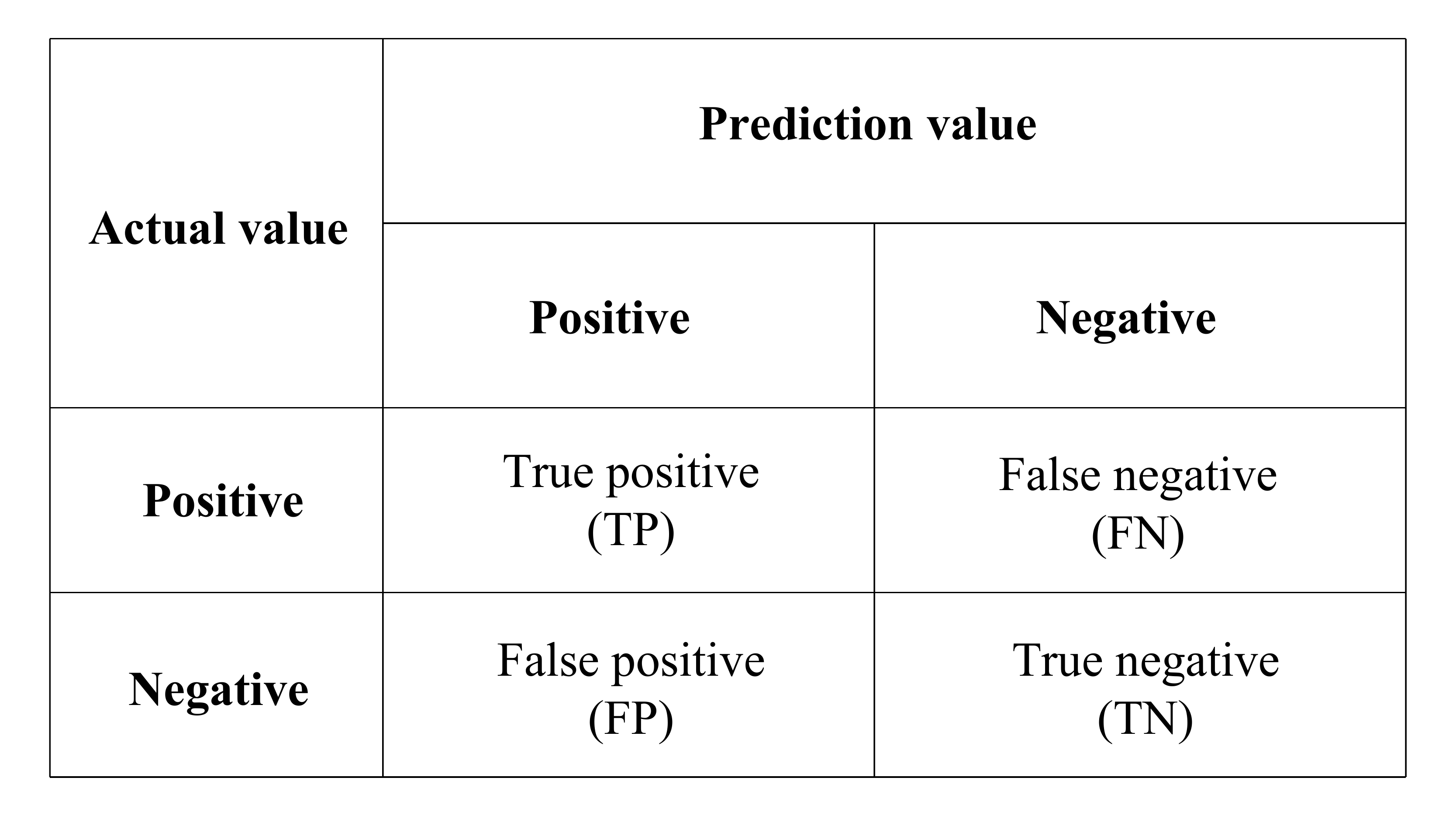}
\caption{Confusion matrix.}
\label{fig:confusionmatrix}
\end{figure}

\textbf{Metrics}. We evaluate the performance of DP-FLMD in terms of both runtime and data utility. To evaluate the data utility of the algorithm, we use F1 score as a measure. A higher F1 score indicates the better utility of the algorithm. Before introducing F1 score, we need to understand the confusion matrix, as shown in Fig. \ref{fig:confusionmatrix}. F1 score can be expressed by Equation \eqref{eq9}, where precision indicates the ratio of TP to FP + TP and recall is expressed as the ratio of TP to FN + TP. F1 score is a metric that combines both precision and recall. It is not possible to fully assess the merits of a model using only precision or recall. Therefore, the previous studies use the weighted average of precision and recall as an evaluation metric \cite{MUC4}.

\begin{equation}\label{eq9}
	\rm F1 \ score =  \frac{2 *  precision * recall}{precision + recall}.
\end{equation}

\subsection{Data utility analysis}

\begin{figure}[!t]
\centering
\includegraphics[width=2.5in]{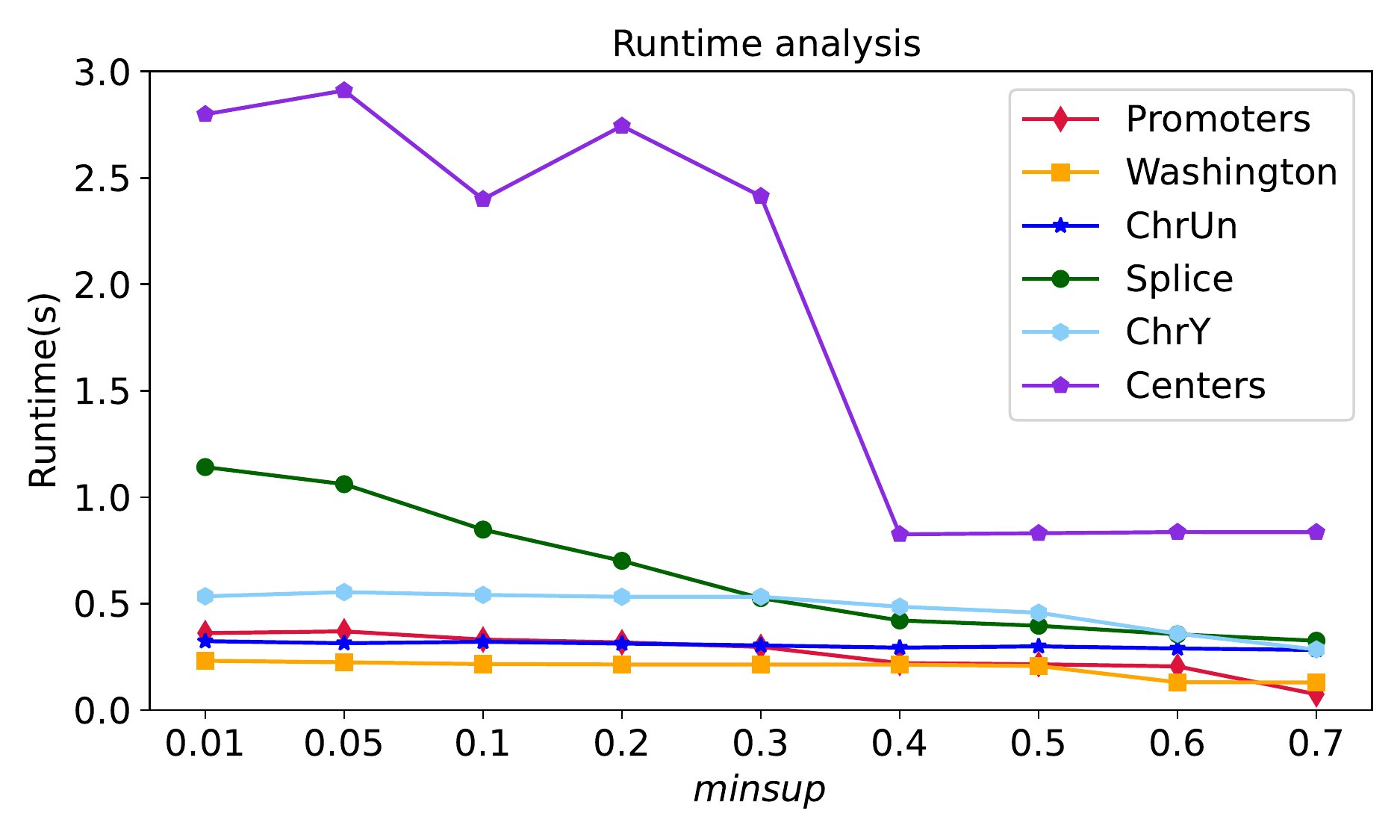}
    \caption{Performance of DP-FLMD in runtime.}
    \label{fig:runtime}
\end{figure}

\begin{figure}[!t]
\centering
\includegraphics[width=2.5in]{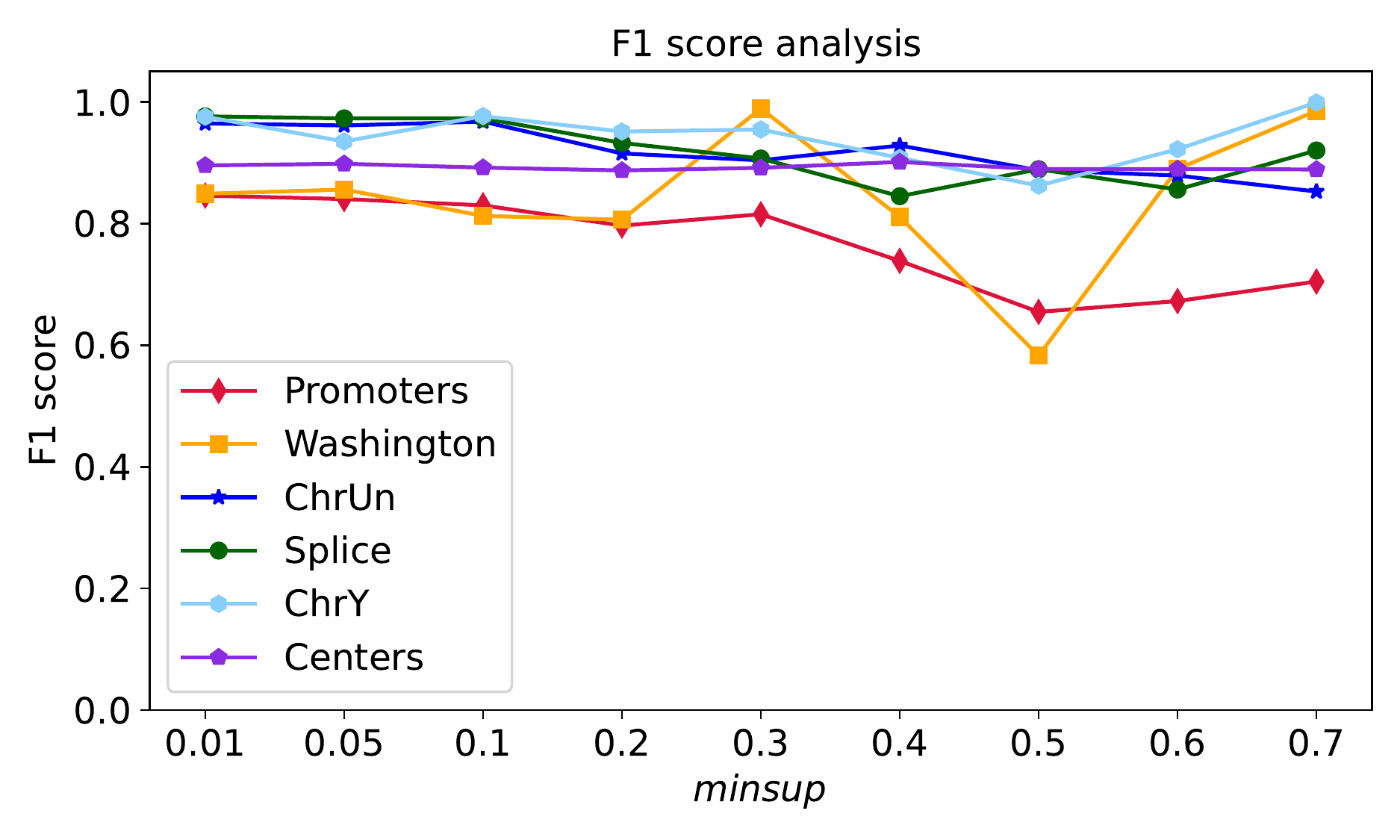}
    \caption{Performance of DP-FLMD in F1 score.}
    \label{fig:f1score}
\end{figure}

We use runtime and F1 score as evaluation metrics. From Fig. \ref{fig:runtime}, it can be seen that the runtime of the dataset Centers is much longer compared to the runtime of the other five datasets. This is because the runtime of DP-FLMD is related to the characteristics of the dataset, such as the number of transactions ($|\mathcal{D}|$) and the average length of a transaction (\textit{avelen}). In addition, the runtime is also related to the number of participants in each round ($x$). Table \ref{number} shows that Centers has higher values of $|\mathcal{D}|$, \textit{avelen}, and $x$ compared to the other five datasets. This explains why the Center has the longest runtime. Promoters, Washington, and ChrUn can be considered as small datasets. From Fig. \ref{fig:runtime}, it can be seen that the difference in runtime among these three small datasets is very small. Their average runtime are 0.266, 0.198, and 0.304, respectively. We can know that Washington has the shortest runtime. This is because Washington has the shortest transaction length (\textit{avelen}). It is well known that the larger the value of \textit{avelen} in a dataset, the longer it takes for the algorithm to mine the patterns. This experimental result verifies the impact of \textit{avelen} on runtime. Fig. \ref{fig:f1score} shows F1 score of DP-FLMD on the six datasets with different parameters. Based on the experimental results, we can see that DP-FLMD has good applicability. In particular, the value of F1 score of DP-FLMD is more stable when dealing with larger datasets.

\subsection{Efficacy of filtering error threshold}

\begin{figure*}[h]
    \centering{
    \includegraphics[scale=0.28]{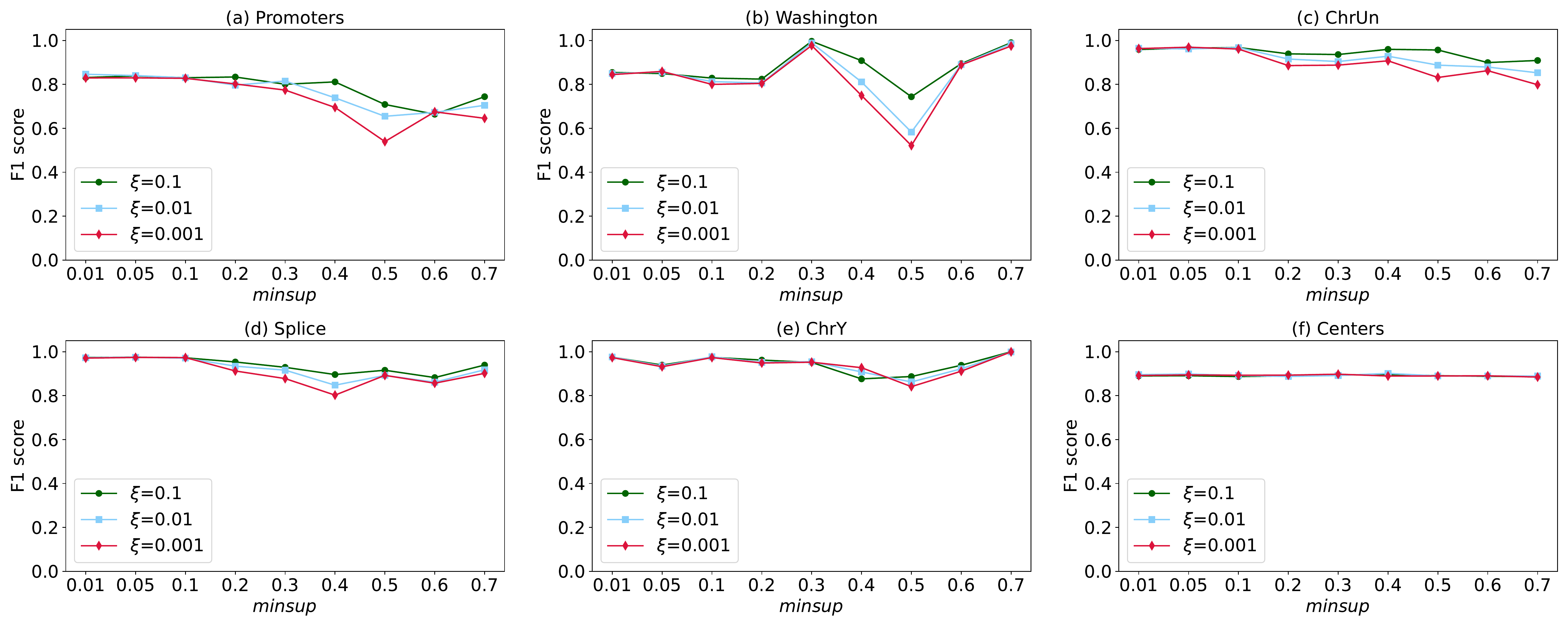}
    }
    \caption{Performance of F1 score of DP-FLMD with different $\xi$.}
    \label{fig:xi}
\end{figure*}

The parameter of the filtering error threshold is denoted as $\xi$. From Fig. \ref{fig:xi}, we find two interesting phenomena. The first phenomenon is that when $\xi$ changes, F1 score of DP-FLMD generally changes as well. We explain this phenomenon using Formula \eqref{eq7} for calculating the modified threshold  (denoted as $\theta$). From Formula \eqref{eq7}, we know that $\xi$ affects the value of $\theta$. When $\xi$ is small, $\theta$ is larger. $\theta$ is used to determine whether the candidate pattern is a frequent pattern or not. This means that $\theta$ affects the mining result of DP-FLMD. Therefore, $\xi$ also affects the mining result of the algorithm. This conclusion is reflected in the fact that F1 scores are generally smaller when $\xi$ is small. The second phenomenon is that the degree of influence of $\xi$ on F1 score is related to the number of transactions ($|\mathcal{D}|$) in the dataset. That is, when the value of $|\mathcal{D}|$ is large, the effect of $\xi$ on F1 score is generally smaller. In particular, in Fig. \ref{fig:xi}(f), there is almost no difference in the value of F1 score when the value of $\xi$ is 0.1, 0.01, or 0.001. In the experiments in this paper, the default value of $x$ is $\frac{|\mathcal{D}|}{2}$. Therefore, when $|\mathcal{D}|$ is large, the value of $x$ will also be large. From Formula \eqref{eq7}, we know that when the value of $x$ is large, the effect of $\xi$ on $\theta$ is small. Therefore, the degree of influence of $\xi$ on F1 score is related to the value of $|\mathcal{D}|$.

\subsection{Performance under different privacy level}

\begin{figure*}[h]
    \centering
    \includegraphics[scale=0.28]{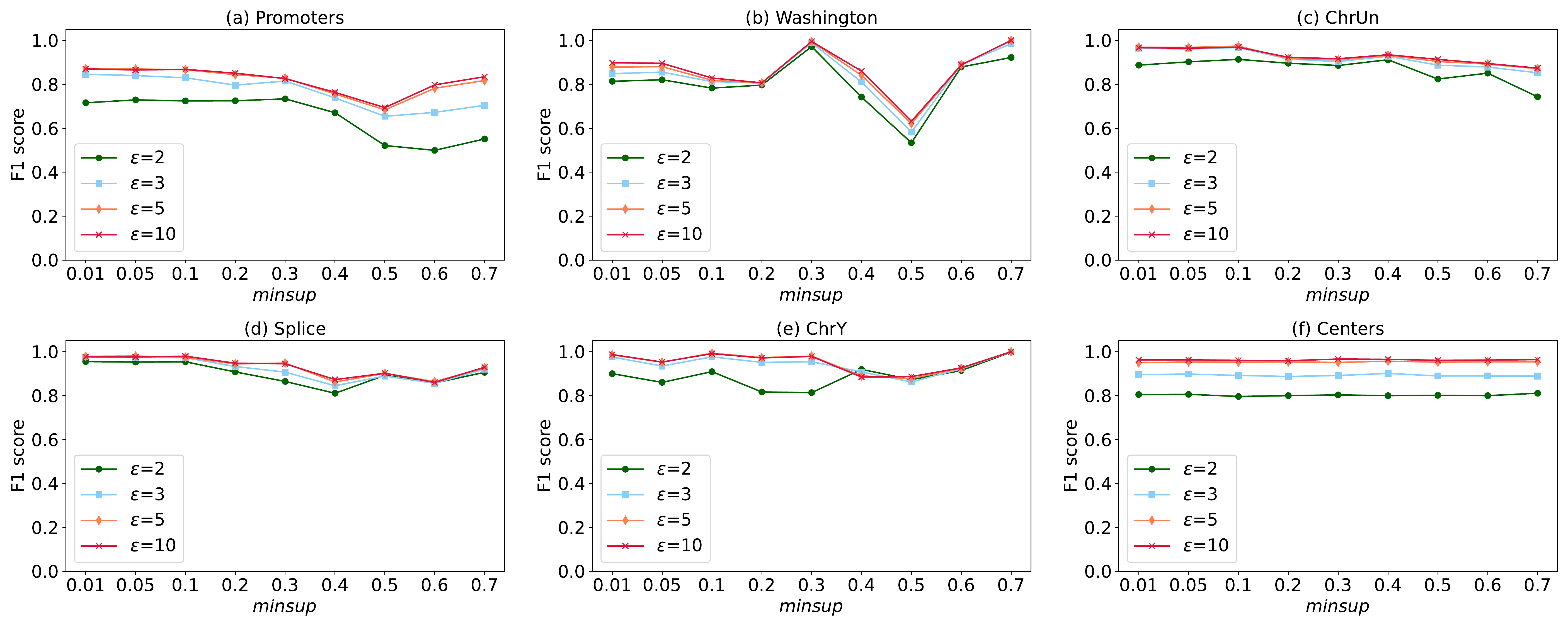}
    \caption{Performance of F1 score of DP-FLMD with different $\epsilon$.}
    \label{fig:epsilon}
\end{figure*}

$\epsilon$ controls the level of privacy protection provided by the randomized response method. A smaller value of $\epsilon$ indicates a higher level of privacy protection. According to Equation \eqref{eq6}, $\epsilon$ is negatively correlated with the noise factor $\eta$. When the value of $\epsilon$ is small, the value of $\eta$ is large. $\eta$ reflects the amount of noise added to the dataset. Therefore, when the value of $\epsilon$ decreases, it increases the level of privacy protection on the one hand and decreases the usability of the algorithm on the other hand. $\epsilon$ quantifies the balance between privacy and usability. To explore its effectiveness, we conducted experiments on different datasets using different values of $\epsilon$. As seen in Fig. \ref{fig:epsilon}, the value of $\epsilon$ affects the performance of the algorithm. When the value of $\epsilon$ is higher, the value of the F1 score is usually higher as well. Therefore, when we want to increase the value of the F1 score, the amount of noise added needs to be reduced (i.e., increase the value of $\epsilon$). In addition, the degree of influence of $\epsilon$ on the F1 score is different if the experimental dataset is different. Because each dataset has different characteristics, including the number of transactions, the average length of a transaction, and data distribution. The results in Fig. \ref{fig:epsilon} validate this conclusion.

In addition, we analyze the relationship between $\epsilon$ and the runtime of the algorithm. In DP-FLMD, regardless of the value of $\epsilon$, each original response performs a randomized response operation and then obtains a noisy response. The server obtains frequent patterns and generates candidate patterns based on the noisy response set. The number of frequent patterns generated in each round affects the number of candidate patterns generated based on the Apriori property, which affects the runtime of the algorithm. Assuming that a lot of noise is added to the dataset (i.e., the value of $\epsilon$ is small). In this case, we can only conclude that the frequent pattern set generated by the algorithm differs significantly from the true frequent pattern set (i.e., the accuracy of the mining results is generally small). It is not possible to conclude that the number of frequent patterns obtained in each round is large or small. Therefore, the relationship between the amount of noise (i.e., the value of $\epsilon$) and the number of frequent patterns generated in each round is neither positive nor negative. This means that it cannot be concluded that there is a monotonic relationship between $\epsilon$ and the runtime of DP-FLMD.

\subsection{Impact of different number of participants}

\begin{figure*}[h]
    \centering
    \includegraphics[scale=0.28]{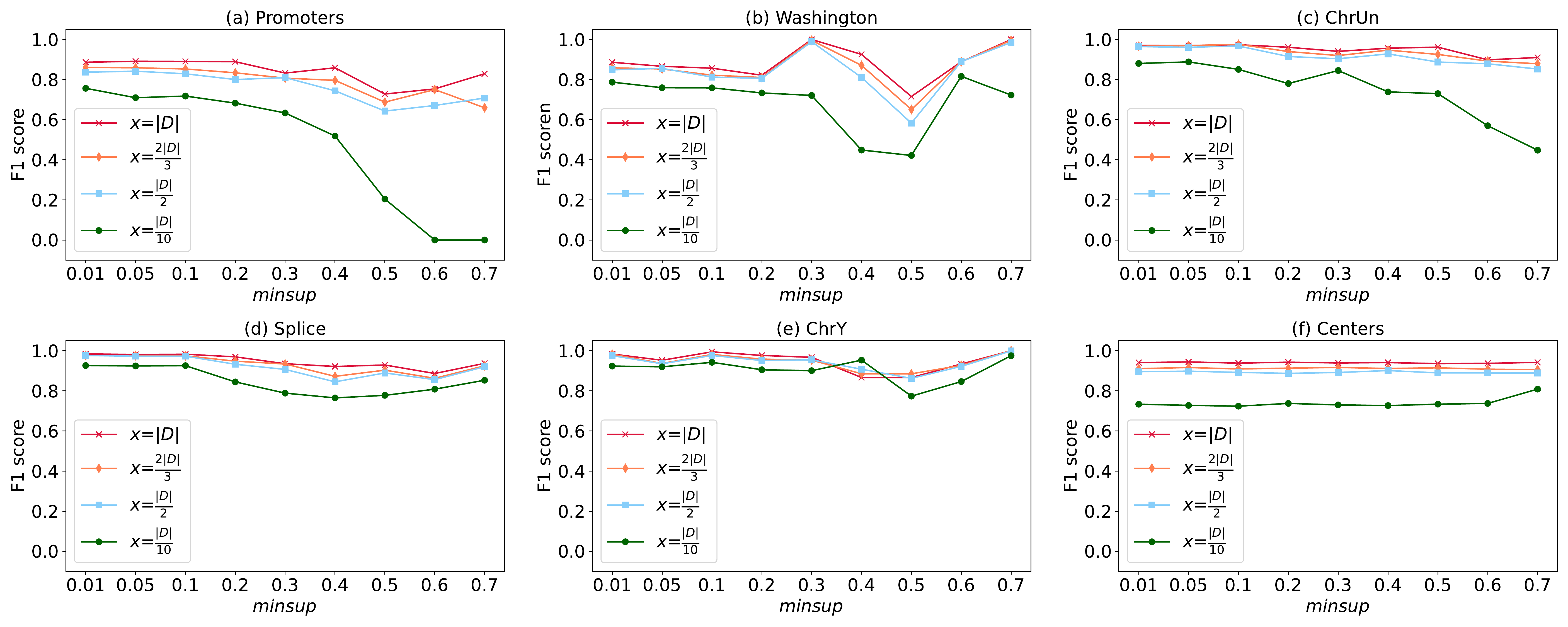}
    \caption{Performance of F1 score of DP-FLMD with different $x$.}
    \label{fig:numparticipants}
\end{figure*}

In real life, sampling methods are often used. Sampling can reduce the amount of work and thus obtain the results faster. In addition, according to probability theory and mathematical statistics, the sampling has a certain probability to ensure the reliability of the results. DP-FLMD selects a certain number of participants from all clients. The number of participants in each round (i.e., the value of $x$)  affects not only the runtime of DP-FLMD, but also the result of the mining. Obviously, when the value of $x$ becomes smaller, the runtime of DP-FLMD becomes shorter. However, a small value of $x$ is likely to lead to unsatisfactory mining results of DP-FLMD. This conclusion can be verified by Fig. \ref{fig:numparticipants}. Since the results of DP-FLMD are randomized, we analyzed the results in the figure from an overall perspective. From Fig. \ref{fig:numparticipants}, we can see that when the value of $x$ is small, the value of F1 score is usually small as well. For small datasets, the value of $x$ is small, which may lead to no pattern being mined (i.e., F1 score is 0). For example, in Fig. \ref{fig:numparticipants}(a), when the support is 0.6 or 0.7 and the value of $x$ is $\frac{|\mathcal{D}|}{10}$,  the number of mined patterns is 0. Therefore, a suitable value of $x$ should be chosen in order to obtain the desired mining results.

\section{Conclusions and Future Work} \label{sec:conclusion}

In order to solve the privacy leakage problem in DNA motif research and alleviate the data silo phenomenon, this paper proposes a novel algorithm for privacy-preserving federated discovery of DNA motifs, named DP-FLMD, based on FL and DP. DP-FLMD uses a query-response method between the server and participants. The server sends sequences to participants for querying. The participants send simple binary answers to respond to the queries from the server, where the binary answers are obtained by the participants executing the LDP method. Then, the server discovers motifs according to the response results of multiple rounds. In addition, in order to reduce the communication cost and total training time, this paper proposes a communication reduction strategy. It reduces the number of messages sent by the server and reduces the overall response time of the participants. Finally, experiments demonstrate that DP-FLMD can obtain highly accurate mining results in a short time and satisfy differential privacy.

In future work, we will improve the DP-FLMD framework in the following three perspectives. To improve the security of the framework, we will continue to learn relevant privacy-preserving techniques and discover whether there are more suitable privacy-preserving methods that can be used in the DNA motif discovery problem. From the perspective of improving the efficiency of the framework, we would like to design more efficient data structures or methods for use in DNA motif discovery problem. From the perspective of improving utility, we would like to apply incentive mechanism to the framework. Because the incentive mechanism helps to increase the interest of participants and reduce the possibility of participants sending useless or harmful data. This may improve the accuracy of the framework. Besides, we will investigate the applicability of this framework, e.g., whether the modified framework can be applied to other pattern mining tasks.

% \appendix
\section*{Appendix A }

According to the definition of $\epsilon-$DP (i.e., Definition \ref{DP}), we know that the algorithm $\mathcal{M}$ satisfies $\epsilon-$DP when Equation \eqref{eq4} holds. Rewriting Equation \eqref{eq4} as:
\begin{equation}\label{eq8}
\forall y \in  Range(\mathcal{M}):  \frac{Pr[\mathcal{M}(d_i) = y]}{Pr[\mathcal{M}(d_j)= y]} \leq e^{\epsilon},
\end{equation}
where $d_i$ and $d_j$ are the possible inputs to the algorithm $\mathcal{M}$.

In the definition of randomized response (i.e., Definition \ref{RR}), we set $\eta \in$ [0, 0.5). Combining with Equation \eqref{eq8}, $\mathcal{M}$ satisfies $\epsilon$-LDP as
\begin{equation}\label{eq10}
	\frac{Pr[\mathcal{M}(d_i) =y]}{Pr[\mathcal{M}(d_j) =y]} \leq \frac{max\{Pr[\mathcal{M}(v) =y]\}}{min\{Pr[\mathcal{M}(v) =y]\}}=  \frac{1 - \eta}{\eta}\leq e^{\epsilon} .
\end{equation}

Using the mathematical derivation, we can conclude that Theorem \ref{theorem:1} is correct.

\section*{Appendix B}

In each round of the federated training, the server sends a candidate pattern (denoted as $p$) to all participants. Suppose there are $x$ participants. Each participant gets an original response by determining whether $p$ exists locally, and then performs a randomized response operation to get a noisy response ($r_n$). The original response and the noisy response ($r_n$) have only  two values, i.e., 0 or 1. Finally, the server collects all noisy responses from the participants. The number of these noisy responses is $x$.

Assume that the number of noisy responses with value 1 is $n_1'$ and the number of noisy responses with value 0 is $n_0'$. Then the average value of $r_n$ is as follows:
\begin{equation}\label{eqB0}
	\overline{r_n} = \frac{n_1' \times 1 + n_0'}{x} = \frac{n_1'}{x}.
\end{equation}

Assume that the number of real responses with value 1 is $n_1$ and the number of real responses with value 0 is $n_0$. Then the true support of $p$ (denoted as $f_t$) is $\frac{n_1}{n_1+n_0}$. $f_t$ also represents the probability that the value of the true response of $p$ is 1.

According to the definition of randomized response (i.e., \ref{RR}), the possibility of the value of the noisy response of $p$ is 1 (denoted as $Pr(r_n=1)$) is expressed as follows:
\begin{equation}\label{eqB1}
    Pr(r_n=1) = f_t(1-\eta) + (1-f_t)\eta = f_t + \eta - 2 \eta f_t.
\end{equation}

According to the definition of mathematical expectation, the expectation of $r_n$ is expressed as:

\begin{equation}\label{eqB2}
\begin{aligned}
\mathbb{E}(r_n) = Pr(r_n = 1) \times 1 + Pr(r_n = 0) \times 0  
\\ 
=  f_t + \eta - 2 \eta f_t =  (1 - 2 \eta )f_t + \eta.
\end{aligned}
\end{equation}

According to Definition \ref{RR}, we know that $\eta$ \textless 0.5. Therefore, $\mathbb{E}(r_n)$ is a monotonic growth function of $f_t$.

The support threshold for frequent patterns given by the user is denoted as $f$. We know that $f_t$ denotes the true support of $p$. Therefore, when $f_t \geq f$, it means that $p$ is a frequent pattern. According to Formula \eqref{eqB2}, we can derive the following formula:
\begin{equation}\label{eqB3}
	f_t \geq f \Leftrightarrow  \mathbb{E}(r_n) \geq  (1-2 \eta)f + \eta,
\end{equation}
where $(1-2 \eta)f + \eta$ is represented by the symbol $\hat{r_n}$, we have
\begin{equation}\label{eqB4}
	\hat{r_n} = (1-2 \eta)f + \eta.
\end{equation}

Hoeffding’s inequality gives a probabilistic bound between the average value of a random variable and  its expected value \cite{hoeffding1994probability}. We apply the Hoeffding’s inequality to the random variable $r_n$. The formula is as follows:
\begin{equation}\label{eqB6}
		\mathbb{P}( \overline{r_n} - \mathbb{E}(r_n) \geq \delta ) \leq exp(-2 \delta^2x).
\end{equation}

Because $\delta$ can be any value, we let $\delta$ satisfy the following equation:
\begin{equation}\label{eqB7}
		exp(-2 \delta^2x) = \xi \Rightarrow \delta = \sqrt{\frac{- ln\xi}{2x}}.
\end{equation}

By substituting Equation \eqref{eqB7} into Equation \eqref{eqB6}, we get
\begin{equation}\label{eqB8}
\mathbb{P}(\overline{r_n} - \sqrt{\frac{- ln\xi}{2x}} \geq \mathbb{E}(r_n)) \leq \xi.
\end{equation}

Therefore, we have
\begin{equation}\label{eqB9}
\mathbb{P}(\overline{r_n} - \sqrt{\frac{- ln\xi}{2x}} \leq \mathbb{E}(r_n)) \geq 1 - \xi.
\end{equation}

We find that $\mathbb{P}(\hat{r_n} \leq \mathbb{E}(r_n)) \geq 1 - \xi $ when the following equation holds.
\begin{equation}\label{eqB10}
		\hat{r_n} \leq \overline{r_n} -  \sqrt{\frac{- ln\xi}{2x}}.
\end{equation}

By substituting Equation \eqref{eqB0} and Equation \eqref{eqB4} into Inequality \eqref{eqB10}, we have
\begin{equation}\label{eqB11}
\frac{n_1'}{x} \geq  f + \eta - 2 \eta f + \sqrt{\frac{- ln\xi}{2x}}.
\end{equation}

Therefore, we can draw a conclusion that when Inequality \eqref{eqB11} is satisfied, $p$ is a frequent pattern with $1 - \xi$ confidence.

% references section
\bibliographystyle{IEEEtran}
% argument is your BibTeX string definitions and bibliography database(s)
\bibliography{DP-FLMD}

\end{document}